\documentstyle[prd,aps]{revtex}
\begin{document}
\draft
\title{Varieties of Dirac equation and flavors of leptons and quarks}
\author{Ruo Peng WANG}
\address{
Physics Department, Peking University, Beijing 100871, P.R.China
}
\date{\today}

\maketitle
%\narrowtext
\begin{abstract}
I show that there exist twelve independent Dirac equations for spin $1/2$ fermions. The Dirac fields that satisfy these equations can be grouped into six pairs according to the way they transform under continuous space-time transformations. These six pairs of Dirac equations correspond to the three quark generations and the three lepton generations. The charged $V-A$ currents can be formed only from fields of the same pair. This property of the Dirac fields implies that a quark or lepton may be transformed only into its partner of the same generation via the charged-current weak interaction. According to the properties of the charged-current weak interaction, I conclude that different elementary fermion fields must satisfy different Dirac equations, and there may not be more than twelve flavors of elementary fermions that are already known. 
\end{abstract}


\section{introduction}

The standard model of the strong and electroweak interactions is used to describe the properties of the quarks and leptons. 
There are six known quarks \cite{herb,Abe,Aba} and six known leptons in nature. The leptons and quarks are spin $1/2$ fermions. They are grouped into three pairs or generations of quarks, and three generations of leptons. Quarks and leptons of same generation can form left-handed isospinors \cite{mandl}, and each member of a quark or lepton generation may be transformed into its partner via the charged-current weak interaction. Within the frame of the standard model, all these elementary fermion fields are assumed to satisfy the Dirac equation of the same form. The only difference between equations for different fermion fields are the fermions' rest masses. Although the standard model has gained already great success, the above assumption is problematic. A quantum field is completely described by its equation and the quantization conditions. If all elementary fermion fields would satisfy the Dirac equation of the same form, then there is not a natural explanation for why a quark or lepton can be transformed only into another specific type of quark or lepton through the charged-current weak interaction, but not into other ones. In other words, there is no natural way to identify which two leptons or quarks should be grouped into a generation, because all leptons and quarks are assumed to satisfy equations of the same type. To be able to describe the behaviors of quarks and leptons under the charged-current weak interaction, we must introduce for each elementary fermion field a different equation. These equations must be grouped into pair in such a way that the vector-minus-axial-vector ($V-A$) weak interaction current can be formed only from the fermion fields satisfying two equations of the same pair. All these equations are for spin $1/2$ massive fields, they must have similar properties as the original Dirac equation, and I call these equations as the varieties of the Dirac equation.

In this paper I will discuss in detail the varieties of the Dirac equation. 
Based on a careful analysis of the properties of the Dirac fields that satisfy these equations, I will try to answer to the questions: why there exist twelve elementary fermions \cite{gail}, and why quarks and leptons appear in three generations. 

\section{Varieties of Dirac equation}

A field of spin $1/2$ transforms under spatial rotation in the following way:

\begin{equation}\label{rtf}
	\psi(x) \longrightarrow \psi^\prime (x^\prime)=e^{i\vec s \cdot \vec  	\theta}\psi(x),
\end{equation}
where $\vec \theta$ is the rotation angle, and $\vec s$ the spin operator that satisfies the following relations:

\begin{equation}\label{spc}
	[s_j,s_k]=i\epsilon_{jkl}s_l,\; s^2_j=\frac{1}{4},\; 
	\;\;\;j,k,l=1,2,3.
\end{equation}
For a four components spinor field, $s_j$ are $4\times 4$ matrices. There exist sixteen linear independent $4\times 4$ matrices, and they can be chosen as

\begin{equation}
	I_4, \;\Sigma^I_j,\;\Sigma^{II}_j,\;\Sigma^I_j\Sigma^{II}_k,\;
	\;\; \;j,k=1,2,3,
\end{equation}
where
\begin{equation}
	I_4=\left(\begin{array}{cc}
	I_2 & 0 \\ 0 & I_2
	\end{array} \right), \;
	\Sigma^{II}_1=\left(\begin{array}{cc}
	0 & I_2 \\ I_2 & 0
	\end{array} \right), \;
	\Sigma^{II}_2=\left(\begin{array}{cc}
	0 & -iI_2 \\ iI_2 & 0
	\end{array} \right), \;
	\Sigma^{II}_3=\left(\begin{array}{cc}
	I_2 & 0 \\ 0 & -I_2
	\end{array} \right),\;
	\Sigma^I_j=\left(\begin{array}{cc}
	\sigma_j & 0 \\ 0 & \sigma_j
	\end{array} \right), \;
	\;j=1,2,3,
\end{equation}
with

\begin{equation}
	I_2=\left(\begin{array}{cc}
	1 & 0 \\ 0 & 1
	\end{array} \right), \;
	\sigma_1=\left(\begin{array}{cc}
	0 & 1 \\ 1 & 0
	\end{array} \right), \;
	\sigma_2=\left(\begin{array}{cc}
	0 & -i \\ i & 0
	\end{array} \right), \;
	\sigma_3=\left(\begin{array}{cc}
	1 & 0 \\ 0 & -1
	\end{array} \right).
\end{equation}
We have

\begin{equation}
	[\Sigma^I_j,\Sigma^{II}_k]=0,\; 
	[\Sigma^I_j,\Sigma^{I}_k]=2i\epsilon_{jkl}\Sigma^{I}_l,\;
	[\Sigma^{II}_j,\Sigma^{II}_k]=2i\epsilon_{jkl}\Sigma^{II}_l,\;
	\;\;\;j,k,l=1,2,3.
\end{equation}
All $4\times 4$ matrices can be written as a linear combination of these sixteen matrices. It easy to verify that there exist only two sets of $4\times 4$ matrices which satisfy the relations (\ref{spc}), these are
\begin{equation}
	s^I_j=\frac{1}{2}\Sigma^I_j,\;s^{II}_j=\frac{1}{2}\Sigma^{II}_j,
	\;\;\;j=1,2,3.
\end{equation}

The Dirac equation for fields of spin $1/2$ and rest mass $m$ is
\cite{drc2}

\begin{equation}\label{drc}
	\Bigl[i \hbar(\frac{\partial}	
	{\partial t} + c \vec {\alpha} \cdot \nabla) 
	-mc^2 \beta \Bigr] \psi(x)=0,
\end{equation}
in which the matrices $\vec {\alpha}$ and $\beta$ satisfy the following 
conditions:

\begin{equation}\label{com1}
	\left.
	\begin{array}{lll}
	\alpha^\dag_{j}=\alpha_{j}, &\alpha^2_{j}=1, 
	& [\alpha_{j},\alpha_{k}]_+=0 \; (j \neq k) \\
	\beta^\dag=\beta, &\beta^2=1, &	[\alpha_{j},\beta]_+=0
	\end{array} \right\}
	j,k=1,2,3,
\end{equation}
where the anticommutator is defined by

\begin{equation}
	[A,B]_+ \equiv AB+BA .
\end{equation}

The matrices $\vec \alpha$ are closely related to the spin operator $\vec s$. The invariance of the Dirac equation under space rotation requests 
\begin{equation}\label{sa}
	[s_j,\alpha_k]=i\epsilon_{jkl}\alpha_l,\;
	[s_j,\beta]=0
	\;\;\;j,k,l=1,2,3.
\end{equation}
Conditions (\ref{com1}) and (\ref{sa}) are satisfied by matrices

\begin{equation}\label{a_b}
	\vec \alpha=\Sigma^I_j \vec \Sigma^{II},\;
	\beta=\Sigma^I_k \;\;\mbox{and}\;\;
	\vec \alpha=\Sigma^{II}_j\vec \Sigma^I, \;
	\beta=\Sigma^{II}_k,\;\;\; k \neq j,\;\; j,k=1,2,3.
\end{equation}
Thus there exist totally twelve varieties of Dirac equation, which should be satisfied by twelve different Fermion fields. 

It's convenient to use the notation
\begin{equation}
	\vec \alpha =2 \lambda \vec s.
\end{equation}
From relations (\ref{com1}) and (\ref{sa}) we get the following conditions for $\lambda$:
\begin{equation}\label{lsb}
	[\lambda, \vec s]=0, \;[\lambda,\beta]_+=0,\;\lambda^2=1.
\end{equation} 
According to Eqs. (\ref{a_b}) we may have

\begin{equation}
	\lambda=\Sigma^{II}_j,\;\beta=\Sigma^{II}_k \;\;\;j,k=1,2,3,\;
	\;\;j\neq k\; \mbox{for}\; \vec s = \vec s^I
\end{equation}
and
\begin{equation}
	\lambda=\Sigma^{I}_j,\;\beta=\Sigma^{I}_k\;\;\;j,k=1,2,3,\;
	\;\;j\neq k\; \mbox{for}\; \vec s = \vec s^{II}.
\end{equation}
One may observe that $\lambda=\Sigma^{II}_1, \vec s=\vec s^I, \beta=\Sigma^{II}_3$ just corresponds to the Dirac-Pauli representation.
The Dirac equation (\ref{drc}) is invariant under Lorentz transformations. A Dirac field transforms under the Lorentz transformation in the following way:

\begin{equation}\label{ltf}
	\psi(x) \longrightarrow \psi^\prime (x^\prime)=
	e^{\frac{1}{2}\vec \alpha \cdot \vec \varphi }\psi(x^\prime)
	= e^{\lambda \vec s \cdot \vec \varphi}\psi(x^\prime).
\end{equation}
By comparing the transformations (\ref{ltf}) and (\ref{rtf}), we find that the operator $\lambda \vec s$ plays a role in the Lorentz transformation similar to the role of $\vec s$ in the space rotation. Therefore we call $\lambda$ as the Lorentz multiplier. The Dirac equation is also invariant under space inversion, and we have under this transformation

\begin{equation}
	\psi(x) \longrightarrow \psi^\prime (x^\prime)=
	e^{i\phi}\beta\psi(x).
\end{equation}
Thus the matrix $\beta$ is nothing but the parity operator, and we find that the twelve varieties of Dirac equation differ among themselves by the spin operator, Lorentz multiplier and the parity operator.

\section{flavors and generations of leptons and quarks}

There are totally twelve varieties of Dirac equations. We can group these equations into six different pairs according to their Lorentz multipliers. 

Let $\psi_1(x)$ and $\psi_2(x)$ be two different fermion fields with the same Lorentz multiplier that satisfy the following equations:

\begin{equation}
	\Bigl[i \hbar(\frac{\partial}	
	{\partial t} + c \Sigma^{II}_1 \vec {\Sigma}^{I} \cdot \nabla) 
	-m_1 c^2 \Sigma_2^{II} \Bigr] \psi_1(x)=0.
\end{equation}
and

\begin{equation}
	\Bigl[i \hbar(\frac{\partial}	
	{\partial t} + c \Sigma^{II}_1 \vec {\Sigma}^{I} \cdot \nabla) 
	-m_2 c^2 \Sigma_3^{II} \Bigr] \psi_2(x)=0.
\end{equation}
The 4-dimensional vector conservation current $j(x)$ of the field $\psi_1(x)$ is defined by

\begin{equation}
	j_{0}(x)=\psi_1^\dag(x)\psi_1(x),\;
	\vec j(x)=\psi^\dag_1(x)\Sigma^{II}_1 \vec {\Sigma}^{I}
	\psi_1(x).
\end{equation}
Under space inversion $\vec j(x)$, the space components of the current $j(x)$, changes to $-\vec j(x^\prime)$. We can also introduce the axial vector current $j^A(x)$ of the field $\psi_1(x)$ defined as

\begin{equation}
	j^A_{0}(x)=\psi_1^\dag(x) \Sigma^{II}_1 \psi_1(x),\;
	\vec j^A(x)=\psi^\dag_1(x) \vec {\Sigma}^{I}
	\psi_1(x).
\end{equation}
The axial vector current $j^A(x)$ transforms as a 4-dimensional vector under continuous space-time transformations. But unlike the current $j(x)$, $ \vec j^A(x)$  remains unchanged under the space inversion.

Because the fermion fields that satisfy Dirac equations with the same Lorentz multiplier transform in the same way under continuous space-time transformation, we may define two quasi-vector currents 

\begin{equation}
	j^q_0=\psi^\dag_1(x)\psi_2(x),\; 
\vec j^q=\psi_1^\dag(x)\Sigma^{II}_1 \vec {\Sigma}^{I}
	\psi_2(x),
\end{equation}
and

\begin{equation}
	j^{qA}_0=\psi^\dag_1(x) \Sigma^{II}_1 \psi_2(x),\; 
\vec j^{qA}=\psi_1^\dag(x) \vec {\Sigma}^{I}
	\psi_2(x),
\end{equation}

It is easy to verify that the quasi-vector currents $j^q(x)$ and $j^{qA}(x)$ transform as 4-dimensional vectors under continuous space-time transformations. But under the space inversion, their behaviors are different from 4-dimensional vectors or axial vectors:

\begin{eqnarray}\label{jqs}
	\vec j^q(x) \longrightarrow  \vec j^{q\prime} (x^\prime)&=&
	e^{i(\phi_2-\phi_1)}\psi_1^\dag(x^\prime)\Sigma^{II}_2	
\Sigma^{II}_1 \vec {\Sigma^{I}} \Sigma^{II}_3\psi_2(x^\prime)
		\nonumber \\
	&=& -i e^{i(\phi_2-\phi_1)}\psi^\dag_1(x^\prime)
	\Sigma^{II}_3 \vec {\Sigma^{I}} \Sigma^{II}_3 \psi_2(x^\prime)
		\nonumber \\
	&=& -i e^{i(\phi_2-\phi_1)}\vec j^{qA}(x)	\nonumber \\
	&\neq& \pm \vec j^q(x^\prime).
\end{eqnarray}
The relation $\Sigma^{II}_2\Sigma^{II}_3=i\Sigma^{II}_1$ was used in (\ref{jqs}). 

We may choose

\begin{equation}
	e^{i(\phi_2-\phi_1)}=-i,
\end{equation}
and we find that the current $J_W(x)$ defined by

\begin{equation}
	J_{W}(x)=j^q(x)-j^{qA}(x)
\end{equation}
transforms just like the $V-A$ current $j(x)-j^A(x)$ under all space-time transformations. In the case of space inversion, we have 

\begin{eqnarray}
	J_{W0}(x) =\psi^\dag_1(x)(1-\Sigma^{II}_1)\psi_2(x)
	&\longrightarrow &
	J_{W0}^\prime (x^\prime)
	=\psi^\dag_1(x^\prime)(1+\Sigma^{II}_1)\psi_2(x^\prime)
	\nonumber \\
\vec J_{W}(x) =-\psi^\dag_1(x)(1-\Sigma^{II}_1)\vec \Sigma^I \psi_2(x^\prime) 
&\longrightarrow &  
\vec J_{W}^\prime (x^\prime)=
	-\psi^\dag_1(x^\prime)(1+\Sigma^{II}_1)\vec \Sigma^I \psi_2(x^\prime).
\end{eqnarray}
 
In fact, $J_W(x)$ is just the $V-A$ charged weak interaction currents. The possibility of forming the $V-A$ charged weak interaction current from two different fermion fields depends on the explicit forms of the Dirac equations for these two fields. The $V-A$ charged weak interaction currents can be formed only from Dirac fields with the same spin operator and the same Lorentz multiplier. By comparing this property of Dirac fields with the behavior of the leptons and the quarks under the charged-current weak interaction, we conclude that the leptons and the quarks of a same generation must satisfy Dirac equations with the same Lorentz multiplier and spin operator. We also conclude that different elementary fermion fields must satisfy different Dirac equations. Because there are twelve varieties of Dirac equations, there must be only twelve different favors of fermions. 

We have shown so far that there exist twelve varieties of Dirac equations that define twelve different fermion fields. The matrices that define these twelve equations can transform among them under certain unitary transformations.

Let
\begin{equation}
	U=\left(\begin{array}{cccc}
	1 & 0 & 0 & 0\\
0 & 0 & 1 & 0\\
0 & 1 & 0 & 0\\
0 & 0 & 0 & 1
	\end{array} \right), 
\end{equation}
then we have
\begin{equation}
	\Sigma^{II}_j=U^\dag \Sigma^I_j U,\;
	\Sigma^{I}_j=U^\dag \Sigma^{II}_j U,\;j=1,2,3.
\end{equation}
We have also

\begin{equation}
	\frac{1}{\sqrt{2}}(1 \mp i\Sigma^{A}_j)\Sigma^{A}_l
	\frac{1}{\sqrt{2}}(1 \pm i\Sigma^{A}_j)
	=\pm \epsilon_{jlk}\Sigma^{A}_k,\;j\neq l, 
	\;\;\;j,l,k=1,2,3;
	\;A=I,II.
\end{equation}

Thus we may transform any of these twelve Dirac equations into another by 
applying a unitary transformation. This unitary transformation can be either $U$, $2^{-1/2}(1 \pm i\Sigma^{I,II}_j)$, or a combination of them.
Due to this symmetry in the forms of Dirac equations, one may choose any type of the Dirac equation to describe a certain lepton or quark field. But when more than one fermion fields are in consideration, one must respect the following rules:
1) fermion fields with different flavors satisfy Dirac equations of different types;
2) lepton or quark fields of the same generation have the same Lorentz multiplier and spin operator, but different parity operator.

For example, if we use the Dirac equation in the Dirac-Pauli representation for electron field $\psi_e(x)$

\begin{equation}
	\Bigl[i \hbar(\frac{\partial}	
	{\partial t} + c \Sigma^{II}_1 \vec {\Sigma}^{I} \cdot \nabla) 
	-m_e c^2 \Sigma_3^{II} \Bigr] \psi_e(x)=0,
\end{equation}
then we must choose the following equation for the electronic neutrino field $\psi_{\nu_e}$:

\begin{equation}
	\Bigl[i \hbar(\frac{\partial}	
	{\partial t} + c \Sigma^{II}_1 \vec {\Sigma}^{I} \cdot \nabla) 
	-m_{\nu_e} c^2 \Sigma_2^{II} \Bigr] \psi_{\nu_e}(x)=0,
\end{equation}
and perhaps the equation

\begin{equation}
	\Bigl[i \hbar(\frac{\partial}	
	{\partial t} + c \Sigma^{II}_2 \vec {\Sigma}^{I} \cdot \nabla) 
	-m_{\mu} c^2 \Sigma_1^{II} \Bigr] \psi_{\mu}(x)=0
\end{equation}
for the muon field $\psi_{\mu}(x)$.

\section{conclusions}

I have shown that there exist twelve independent Dirac equations for spin $1/2$ fermions. Each equation is defined by its spin operator, Lorentz multiplier and parity operator. There are two different spin operators, six different Lorentz multipliers and six different parity operators. These twelve varieties of Dirac equation can be classified into six groups according to the Lorentz multiplier. Each group contains two equations. These six groups of Dirac equations correspond to the three quark generations and the three lepton generations. According to the properties of the charged-current weak interaction, I conclude that different elementary fermion fields must satisfy different Dirac equations, and consequently there may not be more than twelve types of elementary fermions. Or in other words, there can not be elementary fermions other than the already known three generations of quarks and leptons. I have also shown that charged $V-A$ currents can be formed only from fields with the same Lorentz multiplier. This property of Dirac fields implies that quarks and leptons may be transformed via the charged-current weak only into their partner within the same generation.

\end{document}